\documentclass[prl, twocolumn,showpacs,preprintnumbers,amsmath,amssymb]{revtex4}

\usepackage{graphicx}
\usepackage{amsmath}
\usepackage{amssymb}
\usepackage{bm}

\begin{document}

\title{The Effect of Laser Focusing Conditions on Propagation and Monoenergetic Electron Production in Laser Wakefield Accelerators}

\author{A. G. R.~Thomas$^1$}
\author{Z.~Najmudin$^1$}
\author{S. P. D.~Mangles$^1$}
\author{C. D.~Murphy$^{1,2}$}
\author{A. E.~Dangor$^1$}
\author{C. Kamperidis$^{1}$}
\author {K.L. Lancaster$^{2}$}
\author{W. B.~Mori$^3$}
\author{P. A.~Norreys$^2$}
\author{W.~Rozmus$^4$}
\author{K.~Krushelnick$^1$}

\affiliation{$^1$Blackett Laboratory, Imperial College London, SW7 2AZ, UK}
\affiliation{$^2$Central Laser Facility, Rutherford Appleton Laboratory,
Chilton, Oxon, OX11 0QX,  UK} 

\affiliation{$^3$Dept.~of Physics and Dept.~of Electrical Engineering, University of California, Los
Angeles, California 90095} 
\affiliation{$^{4}$Department of Physics, University of Alberta,
Edmonton, Canada}

%\pacs{Valid PACS appear here}

\begin{abstract}
The effect of laser focusing conditions on the evolution of relativistic plasma
waves in laser wakefield accelerators is studied both experimentally and with particle-in-cell simulations. 
For short focal length ($w_0 < \lambda_p$) interactions, beam break-up prevents stable propagation of the pulse. High field gradients lead to non-localized phase injection of electrons, and thus broad energy spread beams. 
However for long focal length geometries ($w_0 > \lambda_p$), a single optical filament can capture the majority of the laser energy, and self-guide over distances comparable to the dephasing length, even for these short-pulses  ($c\tau \approx \lambda_p$). This allows the wakefield to evolve to the correct shape for the production of the monoenergetic electron bunches, as measured in the experiment.
\end{abstract}

\maketitle

The concept of using lasers to accelerate particles to relativistic energies in
plasma \cite{tajima} has been brought closer to practical realization by recent experimental
results. These have demonstrated the production of relativistic electron beams
with high energies \cite{manglesPW, malka}, low energy spread
\cite{mangles,geddes,faure} and low emittance \cite{fritzler}. 
These accelerators promise to revolutionise the many uses of such particle beams, due to their high accelerating fields and thus compact size. Of the numerous
laser based schemes for plasma accelerators \cite{tajima, esarey}, only the
laser wakefield accelerator (LWFA) scheme has been capable of producing quasi-monoenergetic electron beams \cite{mangles, geddes, faure}.
The LWFA consists of a high intensity laser propagating through a plasma,
such that the pulse length is shorter than the relativistic plasma
wavelength ($ c\tau \lesssim \lambda_p = 2\pi c/\omega_p$). 

The longitudinal electric field of the wakefield---a relativistic electron plasma wave, with phase velocity ${\beta_{ph}c}$, generated by the ponderomotive force of the pulse---can be used to accelerate trapped electrons. The maximum electric field the wave can support, in the cold one-dimensional limit, is $E_{max}=[2(\gamma_{ph}-1)]^{1/2} E_0$, where $E_0 = m_{e}c\omega_{p}/e$ and ${ \gamma_{ph} = (1-\beta_{ph}^{2})^{-1/2}}$ \cite{akhiezer}. When
the wave reaches this wavebreaking threshold, the maximum longitudinal velocity of
plasma wave electrons exceeds ${\beta_{ph}c}$. Electrons are self-trapped and accelerated forward, instead of oscillating around their initial position. 

In multiple dimensions the self-trapping mechanism and the wakefield dynamics can be very different from the one-dimensional case \cite{bulanov, Pukhov2,Tsung}. In three dimensions, due to the large radial component of the electron trajectories, wake oscillations are heavily damped \cite{dawson} to the extent that in the extreme case there is only one plasma wave period \cite{Pukhov}. Due to electron cavitation and ion inertia, the electric fields in this wave period are approximately linear and focusing towards the centre of the ion cavity. Electrons slipping back with respect to the pulse are injected into the rear of the cavity where the electric field is strongly forward accelerating. If the charge in the injected bunch is high enough, its own electric field counteracts that of the cavitated region, preventing further injection and creating a short, monoenergetic bunch. 

The electrons in the bunch can reach energies determined by the electric field in the plasma wave and the length over which the acceleration takes place. The maximum energy that an electron can gain is set by the dephasing length $L_D \simeq \gamma_{ph}^2 \lambda_p$, which is the  distance before the relativistic electrons out-run the accelerating phase of the wake \cite{esarey}. For this to happen the laser must propagate at high intensity to the order of this dephasing length, which is generally longer than its Rayleigh length, $z_R$. 

Due to the increased energy requirement and technical difficulty in producing intense, ultra-short pulses, initial LWFA studies used short focal length optics to obtain intensities sufficient to reach the wavebreaking threshold. (A criteria for injection is that the laser normalized vector potential $a_0=eA/m_ec > 4.3$  \cite{Tsung}). However, experimental results with tight focusing have been less successful in terms of charge, electron energy and spectral shape than with longer focal length focusing optics and  $a_0 \approx 1$ \cite{mangles, faure}. 

In this letter we examine the importance of the focusing geometry on the evolution of plasma waves and laser pulse propagation in a LWF accelerator. This is done experimentally by directly comparing laser pulse and electron beam characteristics using two different focusing geometries, chosen to produce either a spot of size $w_0 < \lambda_p$ or $w_0 > \lambda_p$ for $c\tau \lesssim \lambda_p$.  We demonstrate that for $w_0 > \lambda_p$, self-focusing permits $a_0$ to be sufficiently amplified to allow self-injection. Moreover the quasi-static nature of this compression allows for an adiabatic (i.e.~controlled) approach to self-injection, which results in production of monoenergetic bunches.  We also show that the pulse is maintained at high intensity, through self-guiding, over a distance  $\sim L_D$, which is necessary for efficient acceleration of the bunch. With shorter focal length focusing and $w_0 < \lambda_p$, the interaction is characterized by filamented laser propagation and excessive curvature of the plasma wave, resulting in broad electron spectra. The experimental observations are supported by a series of 2D particle-in-cell (PIC) code simulations.

The experiments were carried out on the Ti:Sapphire \textsc{Astra} laser  which provided pulses of energy up to $E_L = 700$ mJ, pulse length $\tau = 50(\pm$5) fs FWHM at a central wavelength
$\lambda_0 = 795$ nm. The laser was focused with either $f/3$ or $f/12$ off-axis parabolic mirrors, where the $f$ refers to the ratio of focal length to initial diameter. The measured spot sizes (FWHM) at focus were $4\:\mu$m and  $20\: \mu$m respectively, resulting in vacuum focused intensities of $5.5\times10^{19}\: \rm{Wcm}^{-2}$ and $2.2\times10^{18}
\:\rm{Wcm}^{-2}$ corresponding to $a_{0}$ of 5 and 1. The $M^2$ of both focal spots was 2.6, giving $z_R$ ($= \pi w_0^2 / M^2 \lambda_0$) of 25 $\mu$m and 600 $\mu$m respectively. 
The laser pulses were focused ($z=0$) onto the front edge of a 3 mm diameter conical helium
gas-jet. This provided initial electron densities of $ 1 \times 10^{18}
\:\rm{cm}^{-3} < \it{n_e} < \rm 3 \times 10^{19} \:\rm{cm}^{-3}$.

The light transmitted through the plasma was measured after reflection by a
glass plate, which served to attenuate the intensity and limit spectral
modifications caused by the window of the target chamber. Light within a $f/5$
cone was collimated and re-imaged onto a 12 bit CCD camera. Light scattered at $90^{\circ}$ to the direction of propagation was reimaged both perpendicular and parallel  to the laser polarization plane.

Electron beam divergence measurements were taken with a removable \textsc{lanex} scintillating screen with a $5^\circ$ opening angle. The energy spectrum of the accelerated electrons was obtained with a magnetic spectrometer with a $5\times 1$ mm steel collimator. The electrons were detected with image plates. These have a linear response over a dynamic range of $10^8$, $50\:\mu$m spatial resolution, and allow calibrated, single-shot acquisition of the full spectrum \cite{imageplates}. Alternatively, for the high charge monoenergetic beams, a \textsc{lanex} scintillating screen was used for high-repetition rate acquisition. 

\begin{figure}[t]
\begin{center}
\includegraphics[width = 3.4in]{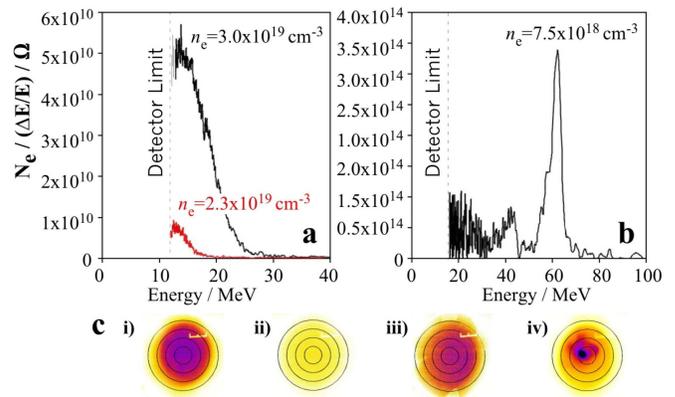}
\caption{(color online) Electron spectra for a) $f/3$ at $n_e = 3.0\times 10^{19} \:\rm{cm^{-3}}$ and
$2.4\times 10^{19}\:\rm{cm^{-3}}$; b) $f/12$  at $n_e = 7.5\times 10^{18}\:\rm{cm^{-3}}$. c) Electron beam divergence measurements for i) $f/3$, $n_e = 3.0\times 10^{19}\:\rm{cm^{-3}}$ ii) $f/3$,  $n_e = 1.5\times 10^{19}\:\rm{cm^{-3}}$ iii) $f/12$, $n_e = 3.0\times 10^{19}\:\rm{cm^{-3}}$ iv) $f/12$, $n_e = 1.5\times 10^{19}\:\rm{cm^{-3}}$. The circles denote $\sim 10 $~mrad intervals} \label{f16spec}
\end{center}
\end{figure}

With both $f/3$ and $f/12$ focusing, at high density ($c\tau > \lambda_{p}$)
quasi-maxwellian electron energy spectra are observed. There is little shot-to-shot variation in spectrum. This is indicative of self-modulation of the pulse \cite{smwf}, where the wakefield amplitude can reach the wavebreaking threshold over multiple wave periods. This, and
the shorter dephasing length causes the large observed energy spread, as in Fig.~\ref{f16spec}a \cite{mangles}. The divergence measurements show an almost uniform beam profile over an instrument limited opening angle of $5^\circ$ as can be seen from Fig.~\ref{f16spec}c i) and iii). Measurements taken with image plates closer to the interaction indicate that the actual divergence is significantly greater than this.

When $c\tau \lesssim  \lambda_p$, the LWF regime is entered. In the case of the $f/3$ the number of accelerated electrons drops dramatically, as can be seen in Fig.~\ref{f16spec}c ii). Coupled with the fall-off in electron flux due to the high beam divergence, the spectrometer can no longer detect any electron signal above noise level, even with the sensitive image plates.

However, with $f/12$ focusing, monoenergetic spectra are observed in the range  $5.6\times 10^{18}\:\rm{cm^{-3}} \leq n_e \leq 2.0\times 10^{19}\:\rm{cm^{-3}}$, Fig.~\ref{f16spec}b. The electron beam divergence monitor consistently measures well collimated beams, with opening half-angle $\lesssim 5\:\rm{mrad}$, Fig.~\ref{f16spec}c iv), in this monoenergetic regime. Large fluctuations in charge are measured by the electron spectrometer. However, this can be explained by the electron beam pointing instability ($3\times 10^{-3} $~sr) being greater than the collection solid-angle of the electron spectrometer collimator  ($5\times 10^{-6} $~sr).

The variation in energy and energy spread may also be affected by this pointing instability. The ratio of the standard deviation to mean of the electron beam peak energy $\sigma / \overline{E} = 0.18$ for a set of 10 consecutive shots at an electron  density of $7.5\times 10^{18}\:\rm{cm^{-3}}$ with $\overline{E} = 80$ MeV. 

\begin{figure}[t]
\begin{center}
\includegraphics[width = 3.2in]{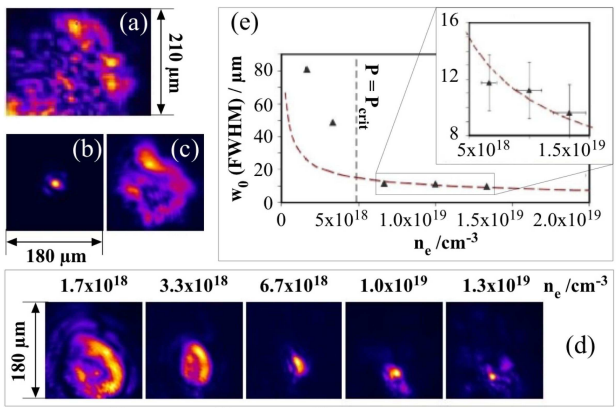}

\caption{(color online) Transmitted light at 800 nm imaged at $z = 2$ mm for (a) $f/3$, $n_e = 1 \times 10^{19}\:\rm{cm^{-3}}$ and (d) $f/12$ at various densities. Also shown is the $f/12$ spot in vacuum (b) at focus and (c) $z = 2$ mm. (e) Mean spot width as a function of $n_e$. Vertical line is $P/P_{cr}=1$, dotted line is $w_0 = \lambda_p$.}
\label{Exitmode}
\end{center}
\end{figure}

Side scattered images of the interaction indicate that the short focal length optic has a short channel length ($\sim z_R$) whilst for the long focal length they indicate extended propagation. Fig.~\ref{Exitmode} shows images of the transmitted light at 800nm at z = 2 mm. For the $f/3$ (Fig.~2a) at all densities within the experimental range, the pulse is broken up into multiple filaments, even when the power for relativistic self-focusing---$P_{cr}=17.3(\omega_0/ \omega_p)^2 \: \rm{GW}$---is exceeded, $P/P_{cr} > 1$. As the density increased the average filament size decreased. 

The lack of guiding indicates why the $f/3$ ($w_0 < \lambda_p$) cannot produce reproducible monoenergetic electrons under these conditions. To produce monoenergetic electron spectra the laser must propagate for distances on the order of the dephasing length. Due to the filamentation, it is unlikely that a single filament will be able to efficiently trap sufficient energy to generate beams of high charge and beam quality \cite{Hidding} compared with those obtained using a long focal length optic.

In contrast, using the $f/12$ optic (Fig.~\ref{Exitmode}d), the pulse emerges as a single filament in the density range of monoenergetic electron spectra ($6.7 - 13\times10^{18}\:\rm{cm^{-3}}$). The spatial extent is comparable to that of the laser focus and scales with $\lambda_p$ (Fig.~2e), indicating that the pulse has self-guided for a distance $\sim 2$ mm. Measurements of the transmitted energy suggest that at this point the pulse has depleted to below a sufficient power to maintain self-guiding. This is in contrast to the $f/3$ case, where almost all of the energy is transmitted. 

For $n_e > 2 \times 10^{19}\:\rm{cm^{-3}}$, the  spatial extent is large again. The depletion is so severe ($> 10 \times$) that self-guiding cannot be maintained over this length. The observed extended propagation allows the necessary non-linear pulse modification for injection of a single monoenergetic bunch of electrons. At the lowest densities ($1.7-3.3\times10^{18}\:\rm{cm^{-3}}$), for which $P/P_{cr}<1$, the spatial extent of the filament is close to that in vacuum at $z=2$ mm. This indicates that self-focusing only partially compensates for diffraction in this case. 
To our knowledge, this is the first evidence for self-guided propagation of short pulses ($c\tau \approx \lambda_p$). Note that previous results that suggested self-focusing is ineffective in this regime, only considered a short focal-length geometry ($w_0 \ll \lambda_p$)  \cite{delfin}.

The two configurations were simulated using the 2D PIC code \textsc{osiris} on an eight node Macintosh G5 cluster. The simulations were run with similar parameters to the experiments, for both the small and large focal spots, at 4 particles/cell using a grid of up to $1500\times 6000$ ($>18$ points/$\lambda$ in $x$ and up to 18 points$/\lambda$ in $y$). A gaussian focus was used, due to the difficulty of initializing a pulse similar to the real focus. The embedded gaussian of the experimental focus was modeled to obtain similar diffraction and Rayleigh lengths. This corresponds to $2~\rm{\mu m}$ and $10~\rm{\mu m}$ beam waists, defined as the half width at $1/e^2$ maximum at vacuum focus.
Simulations with intermediate focusing geometries were also performed. The density profile consisted of a $300\;\rm{\mu m}$ linear ramp followed by a  plateau of density $n_{e0}$ to approximate the measured density profile. The pulse started outside the plasma and its vacuum focus was set to be at the start of the plateau.

\begin{figure}[b]
\begin{center}
\includegraphics[width=3.5in]{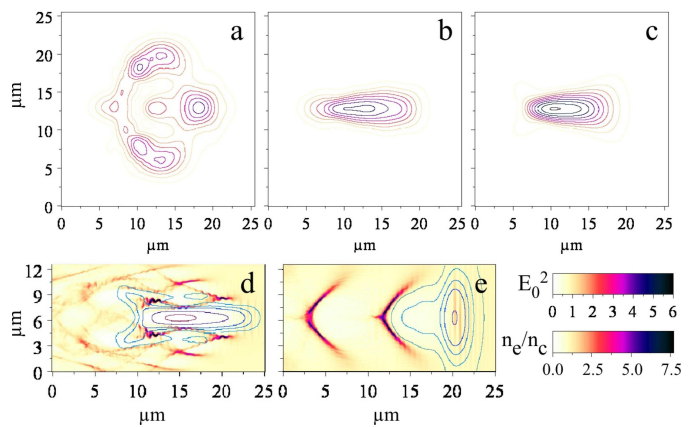}
\caption{(color online) 2D PIC simulations, showing: Contour plots of the laser pulse envelope $(eE/mc\omega_0)^2$ for a) $2~\rm{\mu m}$, b)
$6~\rm{\mu m}$ and c) $10~\rm{\mu m}$ waist pulses after propagating 2$z_R$. $n_e$ plotted for d) $w_0=2~\rm{\mu m}$ pulse at $z=0$,
e) $w_0=10~\rm{\mu m}$ pulse at $z=3z_R$. Overlaid are contours of the pulse at $a=1,1.5,2,3,4$. $E_L$ = 500 mJ, $\tau = 40$ fs,  $n_e = 2 \times10^{19} \rm{cm}^{-3}$.}
\label{simcomp}
\end{center}
\end{figure}

Directly after focus, the $2~\rm{\mu m}$ pulse develops a complex, strongly time-dependent mode structure. This elicits a plasma response that enhances the structure via the self-focusing instability and evolves to a multi-filament structure (Fig.~\ref{simcomp}a) reminiscent of the analytical solution \cite{cattani} in the stationary approximation. Only about 30\% of the pulse energy is trapped in a central filament. 

By contrast the $10~\rm{\mu m}$ pulse focuses radially and compresses temporally, capturing almost all of the initial pulse energy (Fig.~\ref{simcomp}c). 
The average pulse half-width shrinks to $\approx 4\:  \rm{\mu m}$, and the pulse propagates at this width for longer than than $z_R$, until it is depleted. Unlike the $2~\rm{\mu m}$ case, the propagation is quasi-static and self-focusing occurs on a longer timescale.  

Electron density plots are also shown (Fig.~\ref{simcomp}d, e) for both $2~\rm{\mu m}$ and $10~\rm{\mu m}$ simulations at the point where the plasma wave amplitude approaches its maximum value. Note that this amplitude maximum is at very different points in the simulation for the two pulses. Overlaid are contours of the laser $a_0$, showing how the electron density profile directs the laser energy.

The $2~\rm{\mu m}$ pulse interaction is shown close to focus after which it breaks up and the wake amplitude decreases. The electron trajectories from the edge of the pulse are attracted axially by the ion cavity produced by the intense pulse. The distance to reach axis should be similar to the distance the electrons slip back in the speed of light frame in this time, as they are relativistic. However the pulse length in the case of the $f/3$ is much longer than the pulse width, and so the electrons move into the pulse envelope before being repelled by the ponderomotive force of the laser, as can be seen in Fig.~\ref{simcomp}d. It is this crossing of the laser envelope by electrons that causes the break-up of the laser pulse.

The $10~\rm{\mu m}$ pulse produces an initially more gently curved wake and reaches  maximum amplitude later after non-linear modulation has increased its intensity. Even when the intensity is raised so that electron cavitation occurs quickly, there is still no breakup of the pulse as the width is similar to the pulse length.

A scan in waist (between 1 and 14 ${\mu} m$) was simulated at $n_{e0}/n_c = 0.01$,  corresponding to $0.14< w_0/\lambda_{p} < 2$, for a fixed equivalent power, $P=13$~TW. As noted above, larger ($\gtrsim 6~\rm{\mu m}$) focal spots show stable quasi-static propagation (Fig.~\ref{simcomp}b, c), while spot sizes $< 4~\rm{\mu m}$ 
(Fig.~\ref{simcomp}a) show beam break-up and dynamical self-focusing.
Intermediate focusing geometries ($\sim 4~\rm{\mu m}$) show evidence of a transition
region between the two regimes. 

\begin{figure}[b]
\begin{center}
\includegraphics[width=3in]{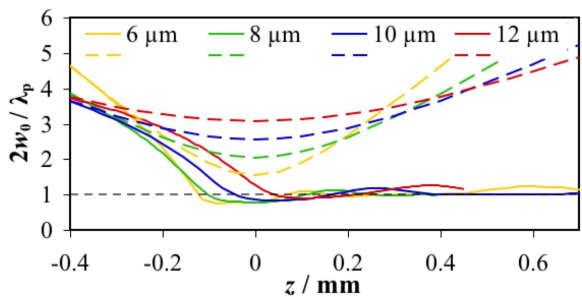}
\caption{(color online) Simulated temporal evolution of spot diameter at $1/e^2$ (2$w$), with various focusing geometries for $n_{e0}/n_{c} = 0.01$, $P/P_{cr} = 2$. Vacuum foci are at $z = 0~\rm{\mu m}$. Solid lines indicate simulated propagation. Dashed lines show theoretical propagation for a gaussian pulse in vacuum.} \label{fstowp}
\end{center}
\end{figure}

An intensity scan ($I_0/4 \leqslant I \leqslant 4 I_0$) at fixed density and a density scan $0.005\leqslant n_e/n_{c} \leqslant 0.02$ at fixed intensity were also simulated for both $2\;\rm{\mu m}$ and $6\;\rm{\mu m}$ geometries. The simulations were qualitatively invariant for both focusing geometries in this parameter range. The simulations also show that the pulse tends to self-focus to a diameter $2w_0\approx \lambda_p$ and oscillate around an equilibrium position (Fig.~\ref{fstowp}). This is independent of the initial $f$ number, provided $w_0~\gtrsim~\lambda_p$. 

Despite the short pulse duration a simple stationary envelope treatment \cite{esarey2} can be used. For $a_0 \gg 1$ the spot size $w$ obeys $\partial^2\overline{w}/\partial^2t = V_0(1/\overline{w}^{3} - 16\sqrt{2}P/P_{cr})$ for a linearly polarized laser. 
Here $V_0 = [(c\lambda_0/\pi)/(w_0^2a_0^2)]^2$, $\overline{w} = w/w_0a_0$. 
Self-focusing will therefore start to dominate diffraction ($\partial^2\overline{w}/\partial t^2 = 0$) for $w/\lambda_p = (\frac{1}{4}P/P_{cr})^{1/6}/\pi$. 
Hence for a mildly relativistic pulse, e.g.~$P/P_{cr} \approx 2$, the matched spot size $2w \approx (2/3)\lambda_p$. 
This is in reasonable agreement with the simulations, the discrepancy being due to the assumptions of constant density and the $a_0 \gg 1$ approximation. For tight focusing geometries and high intensities where $w_0 \lesssim \lambda_p$ and for $a_0 \gg 1$, complete electron blowout can occur \cite{Pukhov} and a non-adiabatic kinetic analysis of self-focusing is required.

In conclusion, we have demonstrated the self-guiding of an ultra-short  laser pulse ($c\tau \sim \lambda_p$) provided the vacuum spot-size $w_0 \gtrsim \lambda_p$. Pulses with vacuum spot-size greater than this can propagate many $z_R$, oscillating about a matched spot size. However, focusing to smaller than $\sim \lambda_p$ will result in mode structures, which through self-focusing seeded beam break-up prevent production of high quality electron beams.  So for reproducible  production of monoenergetic electron, the  ratio of pulse length to spot size should be $\lesssim 1$. To overcome low laser powers, increasing pulse intensity through pulse compression and photon deceleration of the leading edge \cite{Tsung} is  more effective than focusing more tightly. Careful choice of spot size, striking a balance between pump depletion with large $f$ numbers and filamentation at low $f$ numbers, will increase the efficacy of the interaction. This competition between over and under-focusing will be important even at higher intensities and so necessitates a careful choice of focusing conditions in future experiments. 

This work was supported by EPSRC, Alpha-X  and the US DOE. the authors acknowledge the staff of the Central laser Facility (RAL) for technical assistance.

\end{document}